\documentclass[aps,pra,twocolumn,showpacs,groupedaddress]{revtex4}
\usepackage{graphicx}
\usepackage{amsmath}
\usepackage{longtable}

\begin{document}

\title{Positron scattering and annihilation from the hydrogen molecule at zero energy }
\author{J.-Y. Zhang}
\author{J. Mitroy}
%  \email{jxm107@rsphysse.anu.edu.au}
\affiliation{ARC Center for Anti-Matter Studies, School of Engineering, Charles Darwin University, 
Darwin NT 0909, Australia}
\author{K.Varga}
\affiliation{Department of Physics and Astronomy, Vanderbilt University, Nashville, Tennessee 37235, USA}

\date{\today}

\begin{abstract}
 
The confined variational method is used to generate a basis of 
correlated gaussians to describe the interaction region wave function 
for positron scattering from the H$_2$ molecule.  The scattering 
length was $\approx -2.7$ $a_0$ while the zero energy $Z_{\rm eff}$ 
of 15.7 is compatible with experimental values.  The variation of 
the scattering length and $Z_{\rm eff}$ with inter-nuclear distance 
was surprisingly rapid due to virtual state formation at 
$R \approx 3.4$ $a_0$. 
 
\end{abstract}


\maketitle

The lack of spherical symmetry makes the calculation of electron 
or positron scattering from molecules an especially intractable  
computational problem.  The non-spherical potential couples different 
partial waves resulting in an enormous escalation in the size of 
the calculation when compared with atomic targets.  One consequence 
of this is that it is difficult to identify a definitive 
calculation of low energy electron/positron scattering from the 
simplest of molecules, i.e. H$_2$, even under the 
simplifications of the fixed nucleus approximation.      

A new approach to compute the wave function for electron/positron 
scattering from small molecules is developed.  It utilizes existing 
computational technologies from few-body physics that had been used 
to describe the low energy scattering of simple and composite 
projectiles from atoms \cite{zhang08b,mitroy08d,zhang08c}.  
The method is applied to 
the calculation of positron scattering from the H$_2$ molecule.  The cross 
section for positron annihilation at thermal energies was found to be 
compatible with experimental values \cite{mcnutt79a,laricchia87a,wright83a}.  
This is a significant achievement since the annihilation cross section 
presents a stringent test to the accuracy of the scattering wave function 
\cite{armour88a} and its successful prediction solves a previously intractable 
problem.  Our calculations also show the existence of an unexpected virtual 
state at a H$_2$ inter-nuclear distance of $R \approx 3.4$ $a_0$.  

There have been a number of calculations of low energy $e^+$-H$_2$ 
scattering and annihilation  
\cite{hara74a,armour86a,armour88a,sanchez08a,cooper08a}.   
At present, all previous calculations significantly underestimate 
the low energy annihilation cross section.  The most 
sophisticated calculations are the Kohn variational 
calculations performed by Armour and colleagues at the University 
of Nottingham (UN) \cite{armour86a,cooper08a,armour09a}.  Their 
most recent calculations significantly underestimate the annihilation 
cross section at thermal energies.
             
We apply a variant of the confined variational method (CVM)  
\cite{mitroy08d,zhang08b} to describe low energy 
positron-H$_2$ scattering.  In the CVM, an artificial 
confining potential is added to the scattering Hamiltonian thus converting 
the system into a bound system.  This provides a framework that permits 
the wave function in the interaction region to be obtained with bound state 
techniques.  Of crucial importance to this exercise is the use of the 
stochastic variational method (SVM) \cite{kukulin77,suzuki98a,ryzhikh98e} 
to describe the interaction region wave function.  The SVM and variants 
\cite{cencek95a} constitute a powerful tool for studying few 
body systems.  The SVM uses a wave function that is a linear combination 
of explicitly correlated gaussians (ECGs) which have easy to evaluate 
Hamiltonian matrix elements \cite{boys60,suzuki98a}.   
Therefore it is feasible to optimize the non-linear 
parameters of the basis stochastically.   Application to molecular 
systems is easy and ECGs have been recently used to 
describe the wave functions of a number of small molecules to  
high accuracy \cite{cencek08a}.  The close to 
zero energy scattering parameters were extracted from the interaction 
region by a stabilization technique \cite{zhang08b} and 
a technique based on the energy \cite{mitroy08e}.  

The calculation of the interaction region wave function proceeded 
in a manner that was very similar to previous ECG based calculations 
on collision systems \cite{zhang08b,zhang08c}.  The Hamiltonian for 
$e^+\text{H}_2$ scattering was  
\begin{eqnarray}  
H &=& -\sum_{i=0}^2 \frac{\nabla_i^{2}}{2} + \sum_{i=0}^2 W_{\rm CP}(r_i)  
  - \frac{1}{|{\bf r}_0-{\bf r}_1|}  - \frac{1}{|{\bf r}_0-{\bf r}_2|} \nonumber \\  
  &+& \frac{1}{|{\bf r}_1\!-\!{\bf r}_2|} + \frac{1}{|{\bf r}_0-{\bf R}/2|} 
  + \frac{1}{|{\bf r}_0+{\bf R}/2|} - \frac{1}{|{\bf r}_1-{\bf R}/2|} \nonumber \\  
  &-& \frac{1}{|{\bf r}_1+{\bf R}/2|} 
  - \frac{1}{|{\bf r}_2-{\bf R}/2|} - \frac{1}{|{\bf r}_2+{\bf R}/2|} + 
  \frac{1}{R}\ .  
\label{scatHam}
\end{eqnarray} 
The positron coordinate is ${\bf r}_0$ while ${\bf r}_1$ and 
${\bf r}_2$ are the electron coordinates.  The vector ${\bf R}/2$ 
is the displacement of the two protons from the mid-point of 
the molecular axis.  The confining potential $W_{\rm CP}(r)$ 
has the form 
\begin{equation}
W_{\rm CP}(r) = G (r-R_0)^2 \Theta(r-R_0) \ , 
\label{Vform}
\end{equation}
where $\Theta(r-R_0)$ is a Heaviside function and $G$ 
is a small positive number.   

\begin{table*}[bth]
\caption[]{ \label{tab1} 
The convergence of the various properties of the $e^+$-H$_2$ system for the 
$\Sigma_g$ symmetry at $R = 1.4$ $a_0$ as a function of the number of ECGs, $N$.   
The first number in the $N$ column is the dimension of the inner region basis 
while the second entry is the dimension of the outer region basis.  The energy 
of lowest energy state in the confining potential is given by the $E_{N}$ column. 
The wave number, $k$ (in $a_0^{-1}$) is that of the lowest energy pseudo-state 
when the entire basis was diagonalized without the confining potential.  The 
scattering length, $A_{\rm scat}$ (in $a_0$) and $Z_{\rm eff}$ were derived 
from the wave function projections parallel ($\parallel$) and perpendicular 
($\perp$) to the inter-nuclear axis, and from the system energy using the 
soft-box radius (SB).  
 }
\begin{ruledtabular}
\begin{tabular}{lcccccccc}
  \multicolumn{1}{l}{ $N$ } &  \multicolumn{1}{c}{$E_{N}$} &    \multicolumn{1}{c}{$k$ } &  
  \multicolumn{1}{c}{$A_{{\rm scat},\parallel}$}  &  \multicolumn{1}{c}{$A_{{\rm scat},\perp}$}  & 
   \multicolumn{1}{c}{$A_{{\rm scat,SB}}$}  &
   \multicolumn{1}{c}{$Z_{{\rm eff},\parallel}$} & \multicolumn{1}{c}{$Z_{{\rm eff},\perp}$}    
 & \multicolumn{1}{c}{$Z_{{\rm eff,SB}}$} \   \\
\hline
 600+36  & $-$1.16944760  &  0.00635581 &   $-$2.52  &  $-$2.62 & $-$2.59  & 14.38  & 14.48  & 14.41   \\
 800+36  & $-$1.16945780  &  0.00635559 &   $-$2.53  &  $-$2.63 & $-$2.61 & 14.66  & 14.75  & 14.68   \\
1000+36  & $-$1.16946186  &  0.00635551 &   $-$2.53  &  $-$2.63 & $-$2.61  & 14.74  & 14.83  & 14.76  \\
\multicolumn{2}{l}{Kohn: Method of Models, $R = 1.40$ $a_0$, \cite{armour86a}} &  &  \multicolumn{3}{c}{ $-$2.2 } &   \multicolumn{3}{c}{ 10.3 }   \\  
\multicolumn{2}{l}{Kohn: $R = 1.40$ $a_0$, \cite{cooper08a}}       &   &     &    &    &  \multicolumn{3}{c}{ $\approx$ 9.8 }   \\  
\multicolumn{2}{l}{Kohn: Method of Models $R \approx 1.448$ $a_0$, \cite{armour09a}}       &   &     &    &    &  \multicolumn{3}{c}{ $\approx$ 13.5 }   \\  
\multicolumn{3}{l}{Experiment, $k \approx 0.045$ $a_0^{-1}$, $R \approx 1.448$ $a_0$ \cite{mcnutt79a} }  &   &  &    & \multicolumn{3}{c}{ 14.7(2) }   \\
\multicolumn{3}{l}{Experiment,  $k \approx 0.045$ $a_0^{-1}$, $R \approx 1.448$ $a_0$ \cite{laricchia87a}} &   &    &    & \multicolumn{3}{c}{  14.61(14) }   \\
\multicolumn{3}{l}{Experiment,  $k \approx 0.045$ $a_0^{-1}$, $R \approx 1.448$ $a_0$ \cite{wright83a}}    &     &    &    & \multicolumn{3}{c}{  16.02(08) }   \\
\end{tabular}
\end{ruledtabular}
\end{table*}

The first stage of the diagonalization of Eq.~(\ref{scatHam}) was to use the 
SVM to generate an interaction region basis of energy optimized ECGs.  
The ECGs were a generalization of those used previously in 
purely atomic calculations \cite{cencek95a}.  
Their functional form was  
\begin{eqnarray}
\phi_k &=& {\hat P} \ \ \exp \left( -\frac{1}{2}  \sum^{2}_{i=0}b_{k,ij}|{\bf r}_i - {\bf S}_{k,i}|^2 \right) \nonumber \\ 
    & \times & \exp \left( -\frac{1}{2} \sum^{1}_{i=0} \sum^{2}_{j=i+1} a_{k,ij}|{\bf r}_i - {\bf r}_{j}|^2  \right) \ .  
\end{eqnarray}
The vector ${\bf S}_{k,i}$ displaces the center of the ECG for 
the $i$th particle to a point on the inter-nuclear axis.  This 
ensures the 3-particle wave function is of $\Sigma$ symmetry.  
The values of $a_{k,ij}$, $b_{k,ij}$ and ${\bf S}_{k,i}$ are 
adjusted during the optimization process.  The operator 
${\hat P}$ is used to enforce $\Sigma_g$ symmetry.  Each ECG has 
a total of nine stochastically adjustable parameters.  

Table \ref{tab1} lists the energy of the confined $e^+$-H$_2$ system 
for a succession of basis sets.  These energies were generated 
with the confining potential parameters $G = 1.55 \times 10^{-4}$ and 
$R_0 = 18.0$ $a_0$. The inter-nuclear separation was set to 1.40 $a_0$ 
which is very close to the position of the minima in the H$_2$ 
potential curve. 

Extracting scattering information requires embedding the interaction 
region wave function into a formalism for $e^{\pm}$-H$_2$ scattering.  
However, one of our major aims is to demonstrate 
that ECG technologies make it easy to get a good 
description of the $e^+$-H$_2$ collision dynamics.  Accordingly,  
attention is focussed on the very low energy region where the 
outgoing wave is essentially spherical.  

There are two advantages to restricting the current calculation to  
very low energy.  First, the most reliable experimental information 
comes from traditional positron annihilation experiments 
using thermal positrons that yield 
annihilation cross sections at very low energies \cite{charlton85}.  
Second, the collision can be treated as $s$-wave 
scattering and thus the molecular aspects of the asymptotic wave 
function can be neglected with minimal error.   

Positron annihilation cross sections are reported as 
$Z_{\rm eff}$, which is interpreted as the number of electrons 
available for annihilation.  The annihilation cross 
section and $Z_{\rm eff}$ are related by the identity 
\begin{equation}    
Z_{\rm eff}(k) = \frac{k c^3 \sigma_{\rm ann}(k) }{\pi} \ , 
\label{Zeff} 
\end{equation}    
where $c$ is the speed of light.  In the first Born approximation, 
the number of electrons available for annihilation is equal to the 
number of electrons in the molecule.   

The scattering length and near zero energy $Z_{\rm eff}$ were extracted from 
the wave function using a stabilization technique \cite{zhang08b}.  
Initially, the energy optimized interaction region ECG basis is 
supplemented by a set of basis functions to describe the 
long range part of the $e^+$H$_2$ wave function. The functions were    
\begin{eqnarray}
\Psi_{i,out} \! & = & \psi^{{\rm H}_2}({\bf r}_1,{\bf r}_2) \psi_{i}({\bf r}_0) \nonumber  \\ 
\psi_{i}({\bf r}_0) & = & {\hat P}  \  \exp \left( -\frac{1}{2}  \alpha_{i}r_0^2) \right) \ . 
\label{out}
\end{eqnarray}
The target wave function, $\psi^{\rm H_2}({\bf r}_1,{\bf r}_2)$ is  
represented by a linear combination of ECGs.  A basis of dimension 
of 120 gave an energy of $-$1.17447554 a.u..  The H$_2$ energy at an 
inter-nuclear separation of 1.40 $a_0$ is $-$1.17447571 a.u. \cite{cencek08a}.   
Our wave function recovers 99.996$\%$ of the correlation energy of 
0.04084 Hartree \cite{jeziorski77a}.  The $\psi_{i}({\bf r}_0)$   
are designed to describe the positron at asymptotic distances.  The 
$\alpha_k$ were an even tempered set given by the identity  
$\alpha_j = \alpha_1/T^{j-1}$ with $\alpha_1 = 18.59$ and 
$T = 1.435$.  A total of 36 long range basis functions were added to 
interaction region basis.  

The Hamiltonian was then diagonalized (with the confining potential
omitted) with this augmented basis yielding a set of positive energy 
pseudo-states.  The phase shifts were derived by a least squares fit to 
the overlap of the target and projectile wave functions with the
pseudo-states \cite{zhang08b}.  The overlap function, 
$C({\bf r}_0)$ is defined as
\begin{equation}
C({\bf r}_0)  =   \int \ d^3r_1 \ d^3r_2 \  
 \psi^{{\rm H}_2}({\bf r}_1,{\bf r}_2) \Psi({\bf r}_0,{\bf r}_1,{\bf r}_2) \ .
\label{overlap} 
\end{equation}
The overlap function depends on the distance from the inter-nuclear
midpoint and the angle, $\theta_0$ from the inter-nuclear axis.   Least 
squares fits to $r_0 C(r_0)$ over the finite interval, 
$r_0 \in [R_{\rm 1},R_{\rm 2}]$,
at fixed values of $\theta_0$ were made to the asymptotic form 
$B \sin(kr_0+\delta_0)$.  The radial limits for the fit were chosen as 
$R_{\rm 1} = 18$ $a_0$ and $R_{\rm 2} = 30$ $a_0$.  This procedure 
is reminiscent of an earlier method to determine molecular phase 
shifts using discrete functions \cite{mccurdy76a}. The lowest energy 
pseudo-state was at $k \approx 0.006$ $a_0^{-1}$.  The scattering length was 
extracted from the phase shift using $A_{\rm scat} \approx -\tan(\delta)/k$ 
while $Z_{\rm eff}$ is determined from the normalization constant.  
Table \ref{tab1} gives the scattering length and $Z_{\rm eff}$ for 
the lowest energy pseudo-state extracted for projections parallel and 
perpendicular to the inter-nuclear axis.   

An alternate estimate of the scattering length was made from the energy.  
The evenly tempered asymptotic positron basis was diagonalized for a
zero potential.  This basis can be regarded as defining a soft-sided 
box \cite{mitroy08e}.  The effective radius of this box can be estimated from 
the lowest energy $V = 0$ state, and the radius allows the 
scattering length and $Z_{\rm eff}$ to be determined \cite{mitroy08e}.  
These are designated in Table \ref{tab1} as  $A_{\rm scat,SB}$ and 
$Z_{\rm eff,SB}$.  The methods used to estimate the scattering length 
do not take long range polarization and quadrupole interactions into
account past $r_0 \approx 24$ $a_0$. Subsidiary calculations suggest 
an underestimation of $|A_{\rm scat}|$ by about 5$\%$. 

\begin{figure}[tbh]
\centering{
\includegraphics[width=8.6cm,angle=0]{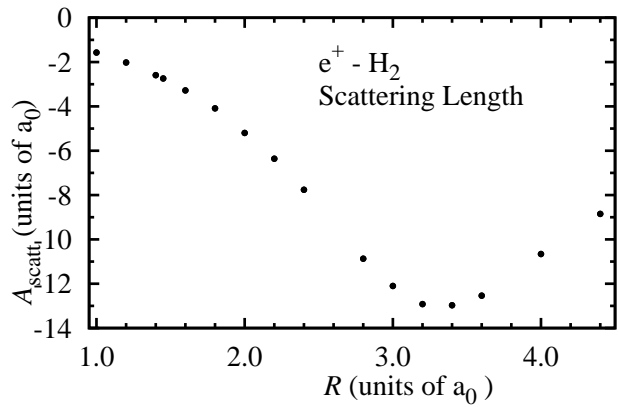}
}
\caption[]{ \label{Ascatt}
The scattering length (in $a_0$) as a function of inter-nuclear
distance, $R$ for positron scattering from H$_2$.
}
\end{figure}

The scattering length in Table \ref{tab1} becomes increasingly negative 
as the dimension of the basis increased.  This is expected on
physical grounds.  Comparison 
between $A_{{\rm scat},\parallel}$, $A_{{\rm scat},\perp}$, and 
$A_{{\rm scat},{\rm SB}}$ and the $Z_{{\rm eff},\parallel,\perp,{\rm SB}}$ 
values reveals the extent to which the low energy scattering parameters are 
largely unaffected by the aspherical potential.  The overall variations 
between the values of $Z_{\rm eff}$ and $A_{\rm scat}$ are about 1$\%$.
The calculations at this energy are equivalent to the H$_2$ molecule 
being its lowest rovibrational level.  It must be kept in mind that 
our calculation is for a fixed axially-symmetric target, while a 
non-Born-Oppenheimer calculation would treat the H$_2$ system as a 
spherically symmetric system.  

The UN group had previously used the method of models within the Kohn 
variational method to determine the low energy $Z_{\rm eff}$.  The 
value listed in Table 1 is taken from the calculations labelled  
``ii'' in Table 4 of \cite{armour86a}.  This gave a $Z_{\rm eff}$ 
of 10.3.  A Kohn variational calculation which explicitly included 
the H$_2$ wave function was very recently reported by the UN group 
\cite{cooper08a}.  The result given in Table \ref{tab1} used a H$_2$ 
wave function which gave 99.7$\%$ of the correlation energy and 
were taken from the $\Psi^{(2,{\rm B})}_{\rm t}$ curves in Figures 
7 and 8 of \cite{cooper08a}.  Some UN method of models calculations 
published while the present letter was under review gave 
$Z_{\rm eff} = 13.5$ \cite{armour09a}.  The same article also gave 
a $Z_{\rm eff} \approx 10$ with an explicit H$_2$ wave function 
and the UN group did not make a clear statement about which result 
should be preferred \cite{armour09a}.  

\begin{figure}[tbh]
\centering{
\includegraphics[width=8.6cm,angle=0]{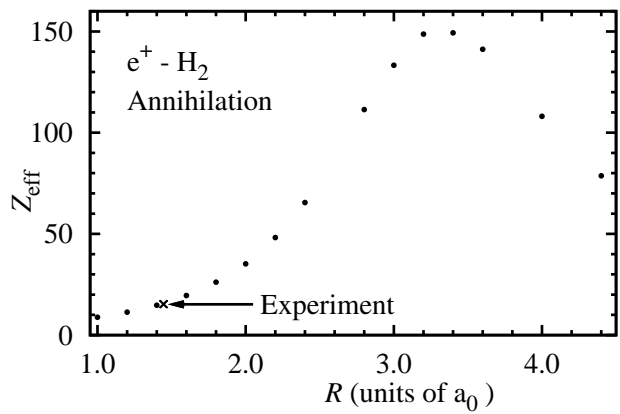}
}
\caption[]{ \label{zeff}
The close to zero energy, $Z_{\rm eff}$, as a function of 
inter-nuclear distance, $R$.  The cross indicates the location 
of the $R = 1.448$ $a_0$ experimental values listed in Table \ref{tab1}.  
}
\end{figure}

Calculations have also been performed at a series of inter-nuclear 
separations between 1.0 $a_0$ and 4.4 $a_0$.  The scattering length as 
a function of inter-nuclear separation is shown in Figure~\ref{Ascatt} 
while the zero-energy $Z_{\rm eff}$ is depicted in Figure \ref{zeff}.    
$Z_{\rm eff}$ for the vibrational ground state was estimated 
by assuming the linear form $Z_{\rm eff}(R) \approx Z_0 + Z_1 R$.  
The $Z_{\rm eff}$ for the vibrational ground state is then   
computed by evaluating $Z_{\rm eff}$ at the mean inter-nuclear 
distance, $\langle R \rangle$. Computing $Z_{\rm eff}$ at 
$\langle R \rangle = 1.448$ $a_0$ \cite{kolos63a} gives 
$\langle Z_{\rm eff} \rangle_{\rm vib} = 15.72$.  The scattering 
length for the vibrational ground state was estimated 
at $-$2.74 $a_0$. 

Experimental $Z_{\rm eff}$ values of 14.7(2) \cite{mcnutt79a}, 
16.02(8) \cite{wright83a} and 14.61(14) \cite{laricchia87a} 
have been measured.  The differences appear to be related to 
variations in $Z_{\rm eff}$ with gas density for reasons that 
are not known \cite{wright83a}.  The present calculation is 
compatible with experiment when consideration is given 
to the uncertainties in the experimental analysis.  
The traditional gas phase positron annihilation experiments  
simply inject high energy positrons into the gas and rely on the 
assumptions that the positrons are thermalized and no other 
processes are occurring when the lifetime spectrum is measured.  

The zero energy vibrational $Z_{\rm eff}$ still needs to be converted 
to thermal energies.  A rough estimate of the size of the correction 
can be made by using an approximate form for the energy dependence 
of $Z_{\rm eff}$ \cite{dzuba96}, e.g.    
\begin{equation}
Z_{\rm eff}(k) = \frac{Z_{\rm eff}(0)}{1+(A_{\rm scat} k)^2} \ . 
\end{equation}
Application of this result with a scattering length of $-$2.7 $a_0$ 
suggests a 1.5$\%$ reduction in the annihilation parameter at 
thermal energies to a value of 15.5. 

The scattering length implies a zero energy cross section of 
$\sigma(0) \approx 30$ $\pi a_0^2$.  A recent experiment by the 
Trento group \cite{zecca09a} had a cross section of 8.3 $\pi a_0^2$ at 
$k \approx 0.086$ $a_0^{-1}$.  The experimental cross section is 
absolutely incompatible with the present scattering length and that 
of the UN group \cite{armour86a}.  Improving the quality of the 
CVM wave function 
would only lead to the magnitude of the scattering length increasing,  
thus leading to larger discrepancies with the Trento cross section 
\cite{zecca09a}.   

The scattering length shows a tendency to increase in magnitude as 
the inter-nuclear separation is increased and a virtual state is
formed around $R \approx 3.4$ $a_0$. The maximum scattering length 
is $-13.0$ $a_0$ at $R = 3.4$ $a_0$.  The peaking of $Z_{\rm eff}$ 
around $3.4$ $a_0$ is expected since it is known that a large 
scattering length leads to a large threshold $Z_{\rm eff}$ \cite{dzuba96}.  
The large scattering length was a surprise.   However it is known that 
the critical value for an electric quadrupole to bind a charged particle 
is 2.4 $ea_0^2$ \cite{ambikaprasad89b}.  The quadrupole moment of H$_2$ 
increases from 0.91 $ea_0^2$ at $R = 1.4$ $a_0$ before reaching a maximum 
value of 2.03 $ea_0^2$ at $R = 3.0$ $a_0$ \cite{poll78a}.  We speculate 
that the large increase in scattering length can be understood in terms of 
the larger quadrupole moment.  The recent method of models calculation 
by the UN group exhibited a qualitatively similar 
variation of $Z_{\rm eff}$ versus $R$ \cite{armour09a}.

While the present calculation was performed under the fixed nucleus 
approximation, it represents the first description with an unrestricted  
treatment of the positron/electron interactions in the $e^+$-H$_2$ collision 
system.  The strong increase in $Z_{\rm eff}$ and $A_{\rm scat}$ with 
increasing inter-nuclear distance due to virtual state formation at 
$R \approx 3.4$ $a_0$ was totally unexpected.  One of the most 
significant methodological aspects was the ease 
with which the inner region wave function was generated.  Using 
the present $e^+$-H$_2$ wave function within a more formal scattering 
framework, such as the Kohn variational method, would require  
substantial development work, but this would involve the application 
of known procedures and would be straightforward.  

This work was supported under the Australian Research Council's
Discovery Program (project number 0665020).  

%\bibliography{positron}

\end{document}